\documentclass[10pt,journal,compsoc]{IEEEtran}
\ifCLASSOPTIONcompsoc
  \usepackage[nocompress]{cite}
\else
  \usepackage{cite}
\fi
\ifCLASSINFOpdf
  \usepackage[pdftex]{graphicx}
\else
\fi
\usepackage[cmex10]{amsmath}
\usepackage{algorithmic}

\usepackage{array}
\usepackage[hyphens]{url}

\begin{document}
%
\title{MLitB: MACHINE LEARNING in the BROWSER}
%
%
%
%

\author{Edward~Meeds,~
        Remco~Hendriks,~
        Said~Al~Faraby,~
        Magiel~Bruntink,~
        and~Max~Welling
\IEEEcompsocitemizethanks{\IEEEcompsocthanksitem All authors are with the Informatics Institute, University of Amsterdam, Amsterdam,
The Netherlands.\protect\\
E-mail: tmeeds@gmail.com
}
}
\IEEEtitleabstractindextext{%
\begin{abstract}
With few exceptions, the field of Machine Learning (ML) research has largely ignored the browser as a computational engine.  Beyond an educational resource for ML, the browser has vast potential to not only improve the state-of-the-art in ML research, but also, inexpensively and on a massive scale, to bring sophisticated  ML learning and prediction to the public at large. This paper introduces MLitB, a prototype ML framework written entirely in JavaScript, capable of performing large-scale distributed computing with heterogeneous classes of devices.  The development of MLitB has been driven by several underlying objectives whose aim is to make ML learning and usage ubiquitous (by using ubiquitous compute devices), cheap and effortlessly distributed, and collaborative.  This is achieved by allowing every internet capable device to run training algorithms and predictive models with no software installation and by saving models in universally readable formats.  Our prototype library is capable of training deep neural networks with synchronized, distributed stochastic gradient descent.  MLitB offers several important opportunities for novel ML research, including: development of distributed learning algorithms, advancement of web GPU algorithms, novel field and mobile applications, privacy preserving computing, and green grid-computing.  MLitB is available as open source software.
\end{abstract}

}

\maketitle

\IEEEdisplaynontitleabstractindextext

%
\IEEEpeerreviewmaketitle

\IEEEraisesectionheading{\section{Introduction}\label{sec:introduction}}

%
%
%
%
\IEEEPARstart{T}{he} field of Machine Learning (ML)  currently lacks a common platform for the development of massively distributed and collaborative computing.   As a result, there are impediments to leveraging and reproducing the work of other ML researchers, potentially slowing down the progress of the field.  The ubiquity of the browser as a computational engine makes it an ideal platform for the development of massively distributed and collaborative ML. Machine Learning in the Browser (MLitB) is an ambitious software development project whose aim is to bring ML, in all its facets, to an audience that includes both the general public and the research community.  

\begin{figure}[!t]
\centering
\includegraphics[width=2.5in]{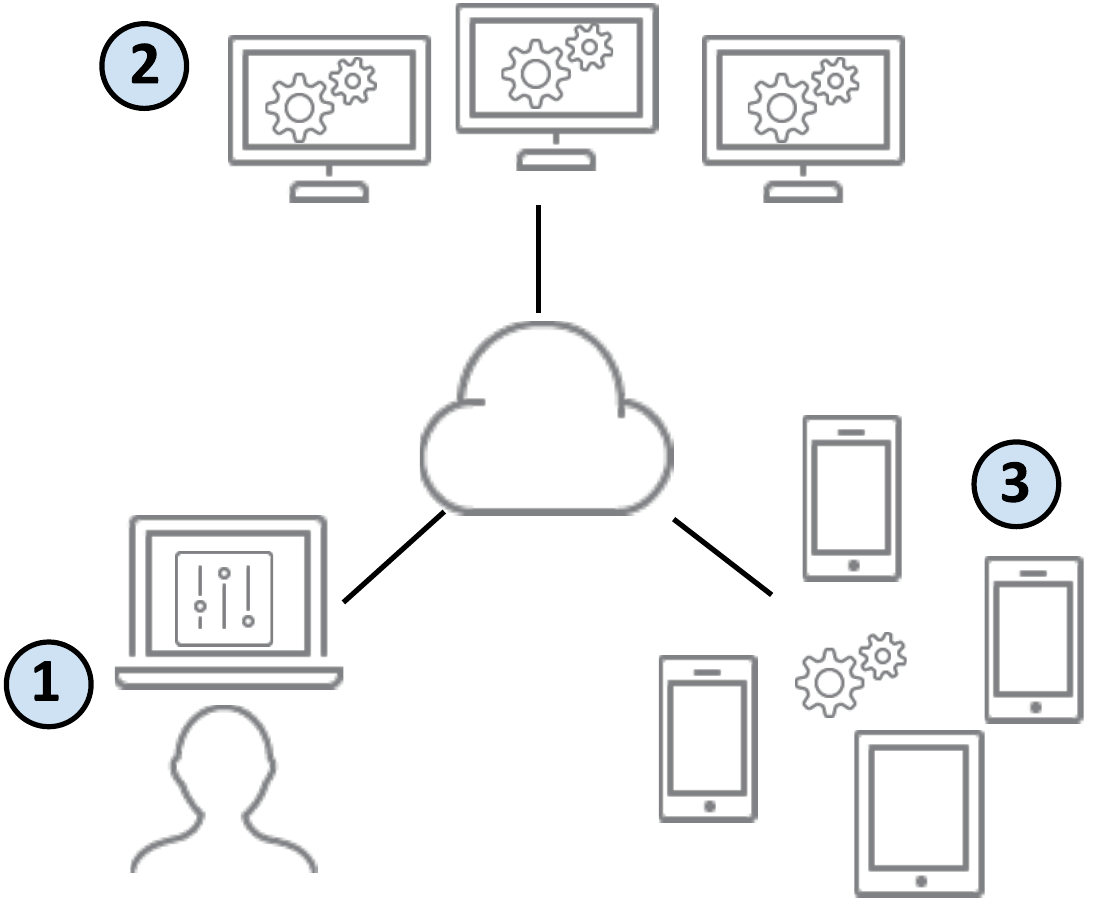}
\caption{{\bf Overview of MLitB.}  {\bf (1)} A researcher sets up a learning problem in his/her browser.  {\bf (2)} Through the internet, grid and desktop machines contribute computation to solve the problem.  {\bf (3)}  Heterogeneous devices, such as mobile phone and tablets, connect to the same problem and contribute computation.  At any time, connected clients can download the model configuration and parameters, or use the model directly in their browsing environment.}
\label{fig:mlitb_overview}
\end{figure}

By writing ML models and algorithms in browser-based programming languages, many research opportunities become available.  The most obvious is software compatibility: nearly all computing devices can collaborate in the training of ML models by contributing some computational resources to the overall training procedure and can, with the same code, harness the power of sophisticated predictive models on the same devices (see Fig.~\ref{fig:mlitb_overview}).  This goal of ubiquitous ML has several important consequences:  training ML models can now occur on a massive, even global scale, with minimal cost, and ML research can now be shared and reproduced everywhere, by everyone, making ML models a freely accessible, public good.  In this paper, we present both a long-term {\bf vision} for MLitB and a light-weight {\bf prototype} implementation of MLitB, that represents a first step in completing the vision, and is based on an important ML use-case, Deep Neural Networks. 
\def \thisscale{0.5}

In Section~\ref{sec:vision} we describe in more detail our vision for MLitB in terms of three main objectives: 1) make ML models and algorithms ubiquitous, for both the public and the scientific community, 2) create an framework for cheap distributed computing by harnessing existing infrastructure and personal devices as novel computing resources, and 3) design {\em research closures}, software objects that archive ML models, algorithms, and parameters to be shared, reused, and in general, support reproducible research.

In Section~\ref{sec:software} we describe the current state of the MLitB software implementation, the MLitB prototype.  We begin with a description of our design choices, including arguments for using JavaScript and the other modern web libraries and utilities.  Then we describe a bespoke map-reduce synchronized event-loop, specifically designed for training a large class of ML models using distributed stochastic gradient descent (SGD).  Our prototype focuses on a specific ML model, Deep Neural Networks (DNNs), using an existing JavaScript implementation \cite{karpathy2014convnetjs}, modified only slightly for MLitB.  We also report results of a scaling experiment, demonstrating the feasibility, but also the engineering challenges of using browsers for distributed ML applications.  We then complete the prototype description with a walk-through of using MLitB to specify and train a neural network for image classification.

MLitB is influenced and inspired by current volunteer computing projects.  These and other related projects, including those from machine learning, are presented in Section~\ref{sec:related}.  Our prototype has exposed several challenges requiring further research and engineering; these are presented in Section~\ref{sec:challenges}, along with discussion of interesting application avenues MLitB makes possible.  The most urgent software development directions follow in Section~\ref{sec:future}.

\section{MLitB: Vision}\label{sec:vision}
Our long-term vision for MLitB is guided by three overarching objectives: 

\noindent{{\bf Ubiquitous ML}:~} models can be training and executed in any web browsing environment without any further software installation.
\\
\noindent{{\bf Cheap distributed computing}:~}  algorithms can be executed on existing grid, cloud, etc., computing resources with minimal (and possibly no) software installation, and can be easily managed remotely via the web; additionally, small internet enabled devices can contribute computational resources.  
\\
\noindent{{\bf Reproducibility}:~} MLitB should foster reproducible science with {\em research closures}, universally readable objects  containing ML model specifications, algorithms, and parameters, that can be used seamlessly to achieve the first two objectives, as well as support sharing of ML models and collaboration within the research community and the public at large.

\subsection{Ubiquitous Machine Learning}
The {\em browser}
is the most ubiquitous computing device of our time, running, in some shape or form on all desktops, laptops, and mobile devices.  Software for state-of-the-art ML algorithms and models, on the other hand, are very sophisticated software libraries written in highly specific programming languages within the ML research community \cite{bastien2012theano,jia2014caffe,collobert2011torch7}. As research tools, these software libraries have been invaluable.  
We argue, however, that to make ML truly ubiquitous requires writing ML models and algorithms with web programming languages and using the browser as the computational engine.  

The software we propose can run sophisticated predictive models on cell phones or super-computers; for the former this extends the distributed nature of ML to a global internet.  By further encapsulating the algorithms and model together, the benefit of powerful predictive modeling becomes a public commodity.

\subsection{Cheap Distributed Computing}   
The usage of web browsers as compute nodes provides the capability of running sophisticated ML algorithms without the expense and technical difficulty of using custom grid or super-computing facilities (e.g. Hadoop cloud computing~\cite{shvachko2010hadoop}).  It has long been a dream to use volunteer computing to achieve real massive scale computing.  Successes include Seti@Home \cite{anderson2002seti} and protein folding \cite{lane2013milliseconds}.   
MLitB is being developed to not only run natively on browsers but also for scaled distributed computing on existing cluster and/or grid resources and, by harnessing the capacity of non-traditional devices, for extremely massive scale computing with a global volunteer base.  
In the former set-up, low communication overhead and homogeneous devices (a ``typical'' grid computing solution) can be exploited.  In the latter, volunteer computing via the internet opens the scaling possibilities tremendously, albeit at the cost of unreliable compute nodes, variable power, limited memory, etc.  Both have serious implications for the user, but, most importantly, both are implemented by the same software.  

Although the current version of MLitB does not provide GPU computing, it does not preclude its implementation in future versions.  It is therefore possible to seamlessly provide GPU computing when available on existing grid computing resources.  Using GPUs on mobile devices is a more delicate proposition since power consumption management is of paramount importance for mobile devices.  However, it is possible for MLitB to manage power intelligently by detecting, for example, if the device is connected to a power source, its temperature, and whether it is actively used for other activities.  A user might volunteer periodic ``mini-bursts'' of GPU power towards a learning problem with minimal disruption to or power consumption from their device.   In other words, MLitB will be able to take advantage of the improvements and breakthroughs of GPU computing for web engines and mobile chips, with minimal software development and/or support. 

\subsection{Reproducible and Collaborative Research}  
Reproducibility is a difficult yet fundamental requirement for science \cite{McNutt14}. 
Reproducibility is now considered just as essential for high-quality research as peer review; simply providing mathematical representations of models and algorithms is no longer considered acceptable~\cite{stodden2013toward}.  Furthermore, merely replicating other work, despite its importance, can be given low publication priority~\cite{casadevall2010reproducible} even though it is considered a prerequisite for publication.  In other words, submissions must demonstrate that their research has been, or could be, independently reproduced. 

For ML research there is no reason for not providing working software that allows reproduction of results (for other fields in science, constraints restricting software publication may exist).  
Currently, the main bottlenecks are the time cost to researchers for making research available, and the incompatibility of the research (i.e. code) for others, which further increases the time investment for researchers.   
One of our primary goals for MLitB is to provide reproducible research with minimal to no time cost to both the primary researcher and other researchers in the community. Following~\cite{Stodden2013Default}, we support ``setting the default to reproducible.''

For ML disciplines, this means other researchers should not only be able to use a model reported in a paper to verify the reported results, but also retrain the model using the reported algorithm.   This higher standard is difficult and time-consuming to achieve, but fortunately this approach is being adopted more and more often, in particular by a sub-discipline of machine learning called {\em deep learning}.  In the deep learning community,  the introduction of new datasets 
 and competitions, along with innovations in algorithms and modeling, have produced a rapid progress on many ML prediction tasks.    Model collections (also called {\em model zoos}), such as those built with  Caffe~\cite{jia2014caffe} make this collaboration explicit and easy to access for researchers.  However, there remains a significant time investment to run any particular deep learning model (these include compilation, library installations, platform dependencies, GPU dependencies, etc).  We argue that these are real barriers to reproducible research and choosing ubiquitous software and compute engines makes it easier.  For example, during our testing we converted a very performant computer vision model~\cite{lin2013network} into JSON  format and it can now be used on any browser with minimal effort.\footnote{JavaScript Object Notation \url{json.org/}}

In a nod to the concept of closures concept common in functional programming, our approach treats a learning problem as a {\em research closure}: a single object containing model and algorithm configuration plus code, along with model parameters that can be executed (and therefore tested and analyzed) by other researchers.  

\section{MLitB: Prototype}\label{sec:software}

The MLitB project and its accompanying software (application programming interfaces (APIs), libraries, etc.) are built entirely in JavaScript. We have taken a pragmatic software development approach to achieve as much of our vision as possible.  To leverage our software development process, we have chosen,  wherever possible, well-supported and actively developed external technology.  By making these choices we have been able to quickly develop a working MLitB prototype that not only satisfies many of our objectives, but is as technologically future proof as possible.  To demonstrate MLitB on a meaningful ML problem, we have similarly incorporated an existing JavaScript implementation of a Deep Neural Network into MLitB.  
  The full implementation of the MLitB prototype can be found on GitHub\footnote{\url{https://github.com/software-engineering-amsterdam/MLitB}}. 

\subsection{Why JavaScript?}\label{sec:why-javascript}

JavaScript is a pervasive web programming language, embedded in approximately $90\%$ of web-sites~\cite{jsusage}. This pervasiveness means it is highly supported~\cite{jssupport}, and is actively developed for efficiency and functionality~\cite{jsv8,jsasm}. As a result, JavaScript is the most popular programming language on GitHub and its popularity is continuing to grow~\cite{Ray:2014}.  

The main challenge for scientific computing with JavaScript is the lack of high-quality scientific libraries compared to platforms such as Matlab and Python.  With the potential of native computational efficiency (or better, GPU computation) becoming available for JavaScript, it is only a matter of time before JavaScript bridges this gap. A recent set of benchmarks showed that numerical JavaScript code can be competitive with native C~\cite{Khan:2014}.

\subsection{General Architecture and Design}\label{sec:softwaregeneral}

\subsubsection*{Design Considerations}
The minimal requirements for MLitB are based on the scenario of running the network as {\em public resource computing}.   The downside of public resource computing is the lack of control over the computing environment.  Participants are free to leave (or join) the network at anytime and their connectivity may be variable with high latency.  MLitB is designed to be robust to these potentially destabilizing events.  The loss of a participant results in the loss of computational power and data allocation.  Most importantly, MLitB must robustly handle new and lost clients, re-allocation of data, and client variability in terms of computational power, storage capacity, and network latency. 

Although we are agnostic to the specific technologies used to fulfill the {\em vision} of MLitB, in practice we are guided by both the requirements of MLitB and our development constraints.  Therefore, as a first step towards implementing our vision, we chose technology pragmatically.  Our choices also follow closely the design principles for web-based big data applications~\cite{Begoli:2012}, which recommend popular standards and light-weight architectures.  As we will see, some of our choices may be limiting at large scale, but they have permitted a successful small-scale MLitB implementation (with up to $100$ clients). 

\begin{figure}[!t]
\centering
\includegraphics[width=2.5in]{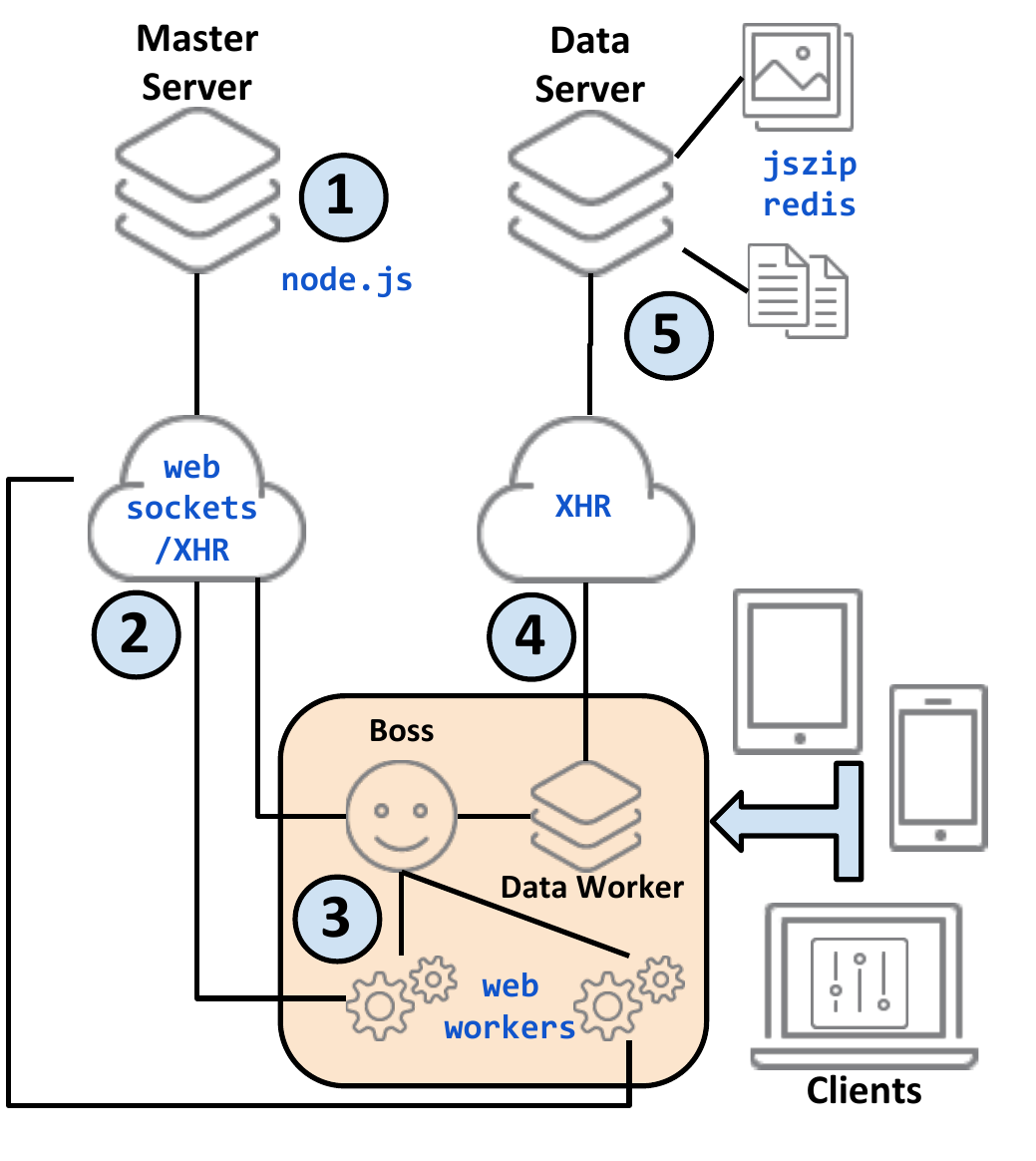}
\caption{{\bf MLitB architecture and technologies.}  {\bf (1)} Servers are {\em Node.js} applications.  The {\em master server} is the main server controlling communication between clients and hosts ML projects.    {\bf (2)} Communication between the master server and clients occurs over Web Sockets. {\bf (3)}  When heterogeneous devices connect to the master server they use Web Workers to perform different tasks.  Upon connection, a UI worker, or boss, is instantiated.  Web Workers perform all the other tasks on a client and are controlled by the boss.  See Fig.~\ref{fig:mlitb_workers} for details. {\bf (4)} A special data worker on the client communicates with the data server using XHR.   {\bf (5)} The {\em data server}, also a Node.js application, manages uploading of data in {\em zip} format and serves data vectors to the client data workers.  
}
\label{fig:mlitb_arch_tech}
\end{figure}

Fig.~\ref{fig:mlitb_arch_tech} shows the high-level architecture and web technologies used in MLitB.  Modern web browsers provide functionality for two essential aspects of MLitB: Web Workers~\cite{webworkers} for parallelizing program execution with threads and Web Sockets~\cite{websockets} for fast bi-directional communication channels to exchange messages more quickly between server and browser.  To maintain compatibility across browser vendors, there is little choice for alternatives to Web Workers and Web Sockets.  These same choices are also used in another browser-based distributed computing platform~\cite{cushing2013distributed}.  

On the server-side, there are many choices that can be made based on scalability, memory management, etc.  However, we chose Node.js for the server application.\footnote{Node.js: \url{http://nodejs.org}.}  Node.js provides several useful features for our application: it is lightweight, written in JavaScript, handles events asynchronously, and can serve many clients concurrently \cite{Tilkov:2010}.
Asynchronous events occur naturally in MLitB as clients join/leave the network, client computations are received by the server, users add new models and otherwise interact with the server.  Since the main computational load is carried by the clients, and not the server, a light-weight server that can handle many clients concurrently is all that is required by MLitB.

%


\subsubsection*{Design Overview}
The general design of MLitB is composed of several parts. A {\em master server} hosts ML problems/projects and connects clients to them.  The master server also manages the {\em main event loop}, where client triggered events are handled, along with the reduce steps of a (bespoke) map-reduce procedure used for computation.    When a browser (i.e. a heterogeneous device) makes an initial connection to the master server, a user-interface (UI) client (aka a {\em boss}) is instantiated.    Through the UI, clients can add {\em workers} that can perform different tasks (e.g., train a model, download parameters, take a picture, etc). An independent {\em data server} serves data to clients using zip files and prevents the master server from blocking while serving data.  For efficiency, data transfer is performed using XHR\footnote{XMLHttpRequest~\url{www.w3.org/TR/XMLHttpRequest}}.  Trained models can be saved into JSON objects at any point in the training process; these can later be loaded in lieu of creating new models.


\def \thisscale{0.5}

  \subsubsection*{Master Server}
  The master node (server) is implemented in Node.js with communication between the master and slave nodes handled by Web Sockets.   The master server hosts multiple ML problems/projects simultaneously along with all clients' connections.  All processes within the master are event-driven, triggered by actions of the slave nodes. Calling the appropriate functions by slave nodes to the master node is handled by the {\em router}.  The master must efficiently perform its tasks (data reallocation and distribution, reduce-steps) because the clients are idle awaiting new parameters before their next work cycle.  New clients must also wait until the end of an iteration before joining a network.  
  The MLitB network is dynamic and permits slave nodes to join and leave during processing.  The master monitors its connections and is able to detect lost participants.  When this occurs, data that was allocated to the lost client is re-allocated the remaining clients, if possible, otherwise it is marked as {\em to be allocated}.

  \subsubsection*{Data Server} 
  The data server is a bespoke application intended to work with our neural network use-case model and can be thought of a lightweight replacement for a proper image database.   The data server is an independent Node.js application that can, but does not necessarily live on the same machine.   Users upload data in zip files before training begins; currently, the data server handles zipped image classification datasets (where sub-directory names define class labels).  
   Data is then downloaded from the data server and zipped files are sent to clients using XHR and unzipped and processed locally.   XHR is used instead of WebSockets  because they communicate large zip-files more efficiently.   A redundant cache of data is stored locally in the clients' browser's memory.  
    For example, a client may store 10,000 data vectors, but at each iteration it may only have the computational power to process 100 data vectors in its scheduled iteration duration.   The data server uses specialized JavaScript APIs {\em unzip.js} and {\em redis-server}.
  
  \subsubsection*{Clients}  
\begin{figure}[!t]
\centering
\includegraphics[width=2.5in]{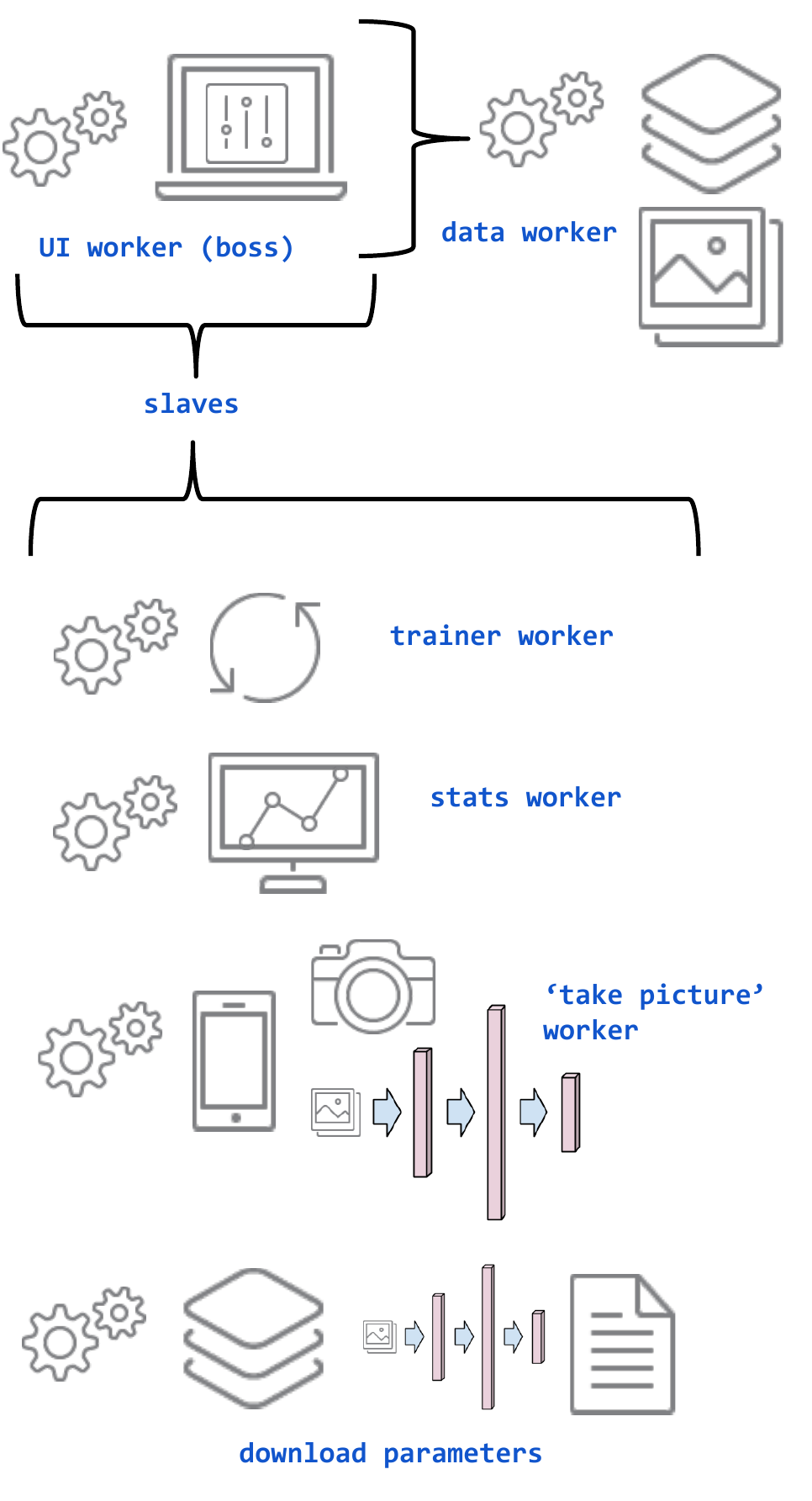}
\caption{{\bf MLitB Client Workers.} Each client connection to the master server initiates a {\em UI worker}, aka a {\em boss}.  For uploading data from a client to the data server and for downloading data from the data server to a client, a separate Web Worker called the {\em data worker} is used.    Users can add slaves through the UI worker; each slave performs a separate task using a Web Worker.}
\label{fig:mlitb_workers}
\end{figure}

  Clients are browser connections from heterogeneous devices that visit the master server's url.  Clients interact through a UI worker, called a {\em boss}, and can create slave workers to perform various tasks (see Workers).  The boss is the main worker running in a client's browser. It manages the slave and image download worker and functions as a bridge between the downloader and slaves. A simple wrapper handles UI interactions, and provides input/output to the boss.  Client bosses use a {\em data worker} to download data from the data server using XHR.  The data worker and server communicate using XHR and pass zip files in both directions.  The boss handles unzipping and decoding data for slaves that request data.   Clients therefore require no software installation other than its native browser.  Clients can contribute to any project hosted by the master server.  Clients can trigger several events through the UI worker.  These include adjusting hyper-parameters, adding data, and adding slave workers, etc. (Fig.~\ref{fig:mlitb_workers}).   Most tasks are run in a separate Web Worker thread (including the boss), ensuring a non-blocking and responsive client UI.  Data downloading is a special task that, via the boss and the data worker, uses XHR to download from the data server.
  
\def \thisscale{0.5}

\subsubsection*{Workers}
 In Fig.~\ref{fig:mlitb_workers} the tasks implemented using Web Worker threads are shown.  At the highest-level is the client UI, with which the user interacts with ML problems and controls their slave workers.  From the client UI, a user can create a new project, load a project from file, upload data to a project, or add slave workers for a project.  Slaves can perform several tasks; most important is the {\em trainer}, which connects to an event loop of a ML project and contributes to its computation (i.e. its map step).  Each slave worker communicates directly to the master server using Web Sockets. For the latter three tasks, the communication is mainly for sending requests for models parameters and receiving them.  The training slave has more complicated behavior because it must download data then perform computation as part of the main event loop.  To train, the user sets the slave task to train and selects start/restart.  This will trigger a join event at the master server; model parameters and data will be downloaded and the slave will begin computation upon completion of the data download. The user can remove a slave at any time.  
 Other slave tasks are {\em tracking}, which requires receiving model parameters from the master, and allows users to monitor statistics of the model on a dataset (e.g. classification error) or to execute the model (e.g. classify an image on a mobile device).  Each slave worker communicates directly to the master server using Web Sockets.
  
\subsection{Events and Software Behavior}

The MLitB network is constructed as a {\em master-slave} relationship, with one server and multiple slave nodes (clients). The setup for computation is similar to a MapReduce network~\cite{dean2008mapreduce}; however, the master server performs many tasks during an {\em iteration} of the {\em master event loop}, including a reduce step, but also several other important tasks.

The specific tasks will be dictated by events triggered by the client, such as requests for parameters, new client workers, removed/lost clients, etc.  Our master event loop can be considered as a synchronized map-reduce algorithm with a user defined iteration duration $T$, where values of $T$ may range from $1$ to $30$ seconds, depending on the size of the network and the problem.  MLitB is not limited to a map-reduce paradigm and in fact we believe that our framework opens the door to peer-to-peer or gossip algorithms~\cite{boyd2006randomized}.  We are currently developing {\em asynchronous} algorithms to improve the scalability of MLitB.

\subsubsection*{Master Event Loop}
The master event loop consists of five steps and is executed by the master server node as long there is at least one slave node connected. Each loop includes one map-reduce step, and runs for at least $T$ seconds. The following steps are executed, in order:

\begin{enumerate}
  \item[a)] New data uploading and allocation.
  \item[b)] New client trainer initialization and data allocation.
  \item[c)] Training workers reduce step.  
  \item[d)] Latency monitoring and data allocation adjustment.
  \item[e)] Master broadcasts parameters.  
\end{enumerate}

\subsubsection*{{\bf a) New data uploading and allocation}}
When a client boss uploads data, it directly communicates with the data server using XHR.  Once the data server has uploaded the zip file, it sends the data indices and classification labels to the boss.  The boss then registers the indices with the master server.  Each data index is managed: MLitB stores an {\em allocated} index (the worker that is allocated the id) and a {\em cached} index (the worker that has cached the id).  The master ensures that the data  allocation is balanced amongst its clients.  Once a data set is allocated on the master server, the master allocates indices and sends the set of ids to workers.  Workers can then request data from the boss, who in turn use its data downloader worker to download those worker specific ids from the data server.  The data server sends a zipped file to the data downloader, which are then unzipped and processed by the boss (e.g. JPEG decoding for images).  The zip file transfers are fast but the decoding can be slow.  We therefore allow workers to begin computing before the entire dataset is downloaded and decoded, allowing projects to start training almost immediately while data gets cached in the background.
  
\subsubsection*{{\bf b) New client trainer initialization and data allocation}}
  When a client boss adds a new slave, a request to join the project is sent to the master.  If there is unallocated data, a balanced fraction of the data is allocated to the new worker.  If there is no unallocated data, a pie-cutter algorithm is used to remove allocated data from other clients and assign it to the new client.  This prevents unnecessary data transfers.  The new worker is sent a set of data ids it will need to download from the client's data worker.  Once the data has been downloaded and put into the new worker's cache, the master will then add the new worker to the computation performed at each iteration.  The master server is immediately informed when a client or one of its workers is removed from the network.\footnote{If a user closes a client tab, the master will know immediately and take action.  In the current implementation, if a user closes the master tab, all current connections are lost.}  Because of this, it can manage the newly unallocated data (that were allocated to the lost client).  

\subsubsection*{{\bf c) Training workers' reduce step}} The reduce step is completely problem specific.  In our prototype, workers compute gradients with respect to model parameters over their allocated data vectors, and the reduce step sums over the gradients and updates the model parameters.
  
\subsubsection*{{\bf d) Latency monitoring and data allocation adjustment}}  
 The interval $T$ represents both the time of computation {\em and} the latency between the client and the master node.    The synchronization is stochastic and adaptive.  At each reduce step, the master node estimates the latency between the client and the master and informs the client worker how long it should run for.  A client does not need to have a batch size because it just clocks its own computation and returns results at the end of its scheduled work time.   
 Under this setting, it is possible to have mobile devices that compute only a few gradients per second and a powerful desktop machine that performs hundreds or thousands.  This simple approach also allows the master to account for unexpected user activity: if the user's device slows or has increased latency, the master will decrease the load on the device for the next iteration.  Generally, devices with a cellular network connection communicate with longer delays than hardwired machines. In practice, this means the reduction step in the master node receives delayed responses from slave nodes, forcing it to run the reduction function after the slowest slave node (with largest latency) has returned. This is called {\em asynchronous reduction callback delay}.
  
\subsubsection*{{\bf e) Master broadcasts parameters}} An array of model parameters is broadcast to each clients' boss worker using XHR; when the boss receives new parameters, they are given to each of its workers who then start another computation iteration.

\subsection{ML use-case: Deep Neural Networks}\label{sec:softwarednns}
The current version of the MLitB software is built around a pervasive ML use-case: deep neural networks (DNNs). DNNs are the current state-of-the-art prediction models for many tasks, including computer vision~\cite{krizhevsky2012imagenet,lin2013network}, speech recognition~\cite{hinton2012deep}, and natural language processing and machine translation~\cite{liu2014recursive,bahdanau2014neural,sutskever2014sequence}.  Our implementation only required {\em superficial} modifications to an existing JavaScript implementation~\cite{karpathy2014convnetjs} to fit into our network design.

\begin{figure}[t!]
\centering
\includegraphics[width=0.9\columnwidth]{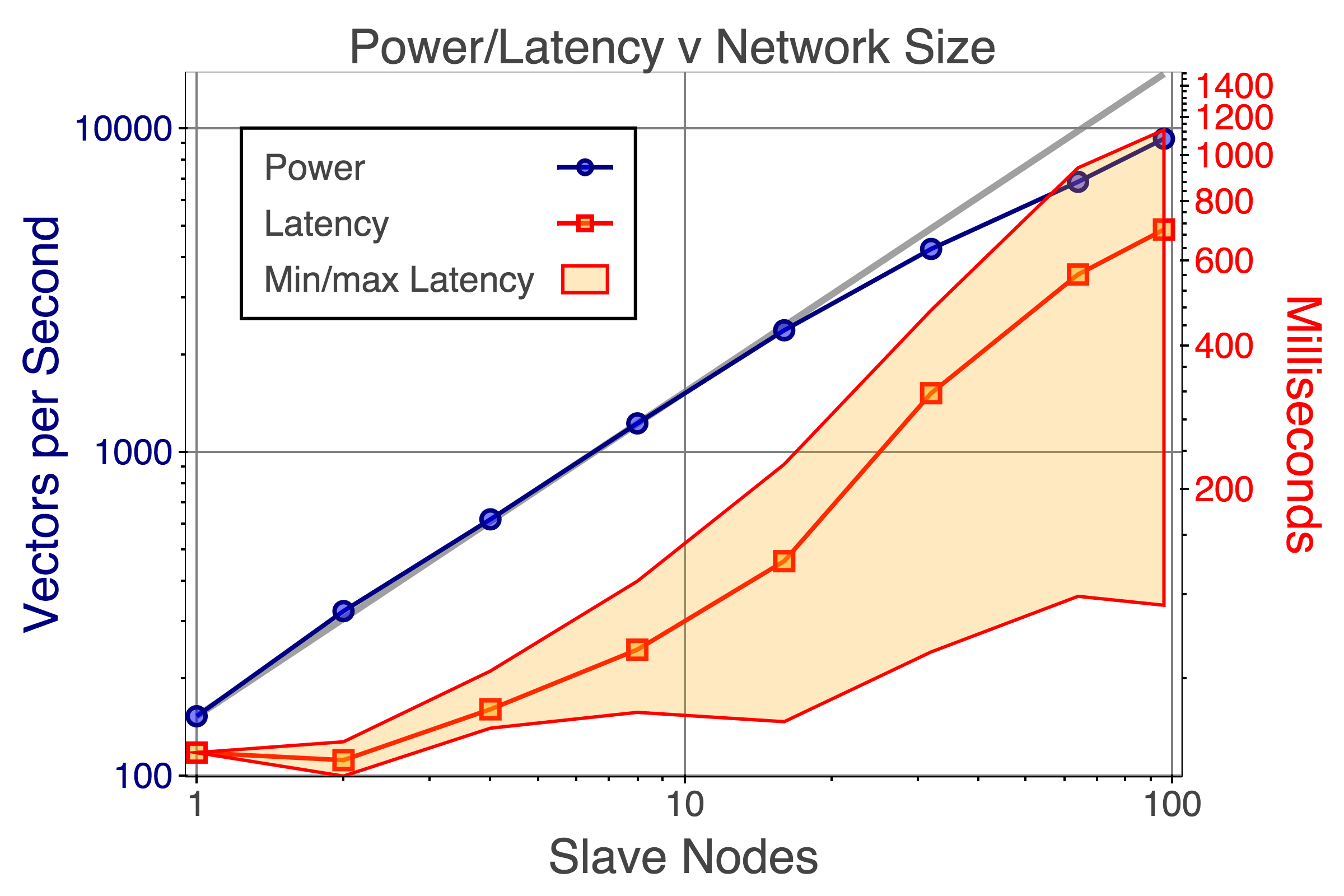}
\caption{{\bf Effects of scaling on power and latency.} Power---measured as the number of data vectors processed per second---scales linearly until 64 nodes, when the increase in latency jumps.  The ideal linear scaling is shown in grey.
}
\label{fig:scaling-experiments-1}
\end{figure}

\subsection{Scaling Behavior of MLitB}\label{sec:scaling-experiment}
We performed an experiment to study the scaling behavior of MLitB prototype.  Using up to 32 4-core workstation machines connected on a local area network using a single router, we trained a simple convolutional NN on the MNIST dataset for 100 iterations (with 4 seconds per iteration/synchronization event).\footnote{Slave node specifications (32 units): Intel Core i3-2120 3.3GHz (dual-core); 4GB RAM; Windows 7 Enterprise x64; Google Chrome 35.  Master node specifications (1 unit): Intel Xeon E5620 2.4GHz (quad-core); 24 GB RAM; Ubuntu 10.04 LTS. NodeJS version: v0.10.28.  The NN has a $28 \times 28$ input layer connected to $16$ convolution filters (with pooling), followed by a fully connected output layer.}  The number of slave nodes doubled from one experiment to the next (i.e. $1, 2, 4, \ldots, 96$).  We are interested in the scaling behavior of two performance indicators: 1) power, measured in data vectors processed per second, and 2) latency in milliseconds between slaves and master node.  Of secondary interest is the generalization performance on the MNIST test set.  As a feasibility study of a distributed ML framework, we are most interested scaling power while minimizing latency effects during training, but we also want to ensure the correctness of the training algorithm.  Since optimization using compiled JS and/or GPUs of the ML JavaScript library possible, but not our focus, we are less concerned with the power performance of a single slave node.

\begin{figure}[b!]
\centering
\includegraphics[width=0.9\columnwidth]{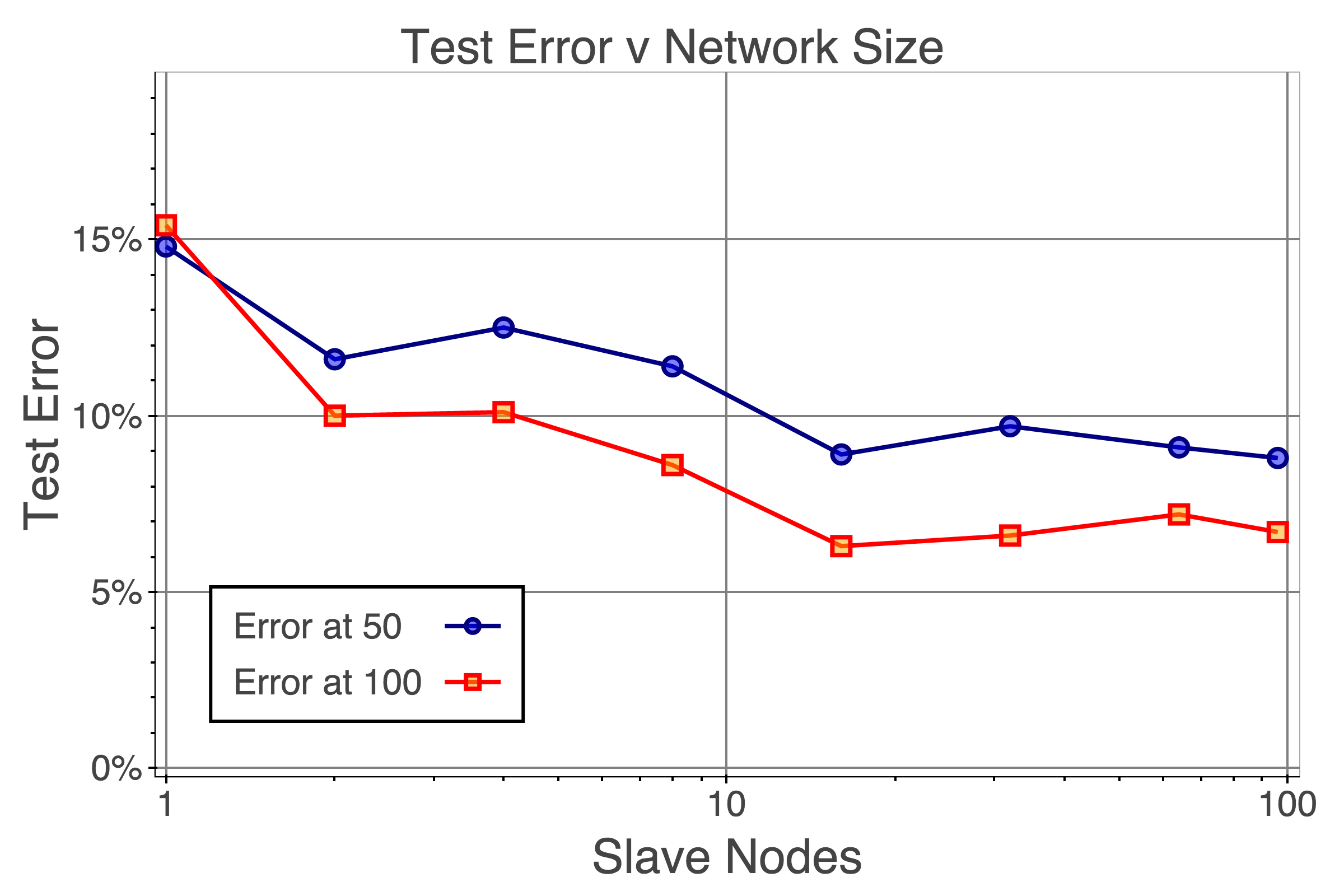}
\caption{{\bf Effects of scaling on optimization.}  Convergence of the NN is measured in terms of test error after 50 and 100 iterations.  Each point represents approximately the same wall-clock time (200/400 seconds for 50 and 100 iterations, respectively).  
}
\label{fig:scaling-experiments-2}
\end{figure}
Results for power and latency are shown in Fig.~\ref{fig:scaling-experiments-1}.  Power increases linearly up to 64 slave nodes, at which point a large increase in latency limits additional power gains for new nodes.  This is due to a single server reaching the limit of its capacity to process incoming gradients synchronously.  Solutions include using multiple server processes, asynchronous updates, and partial gradient communication.  Test error, as a function of the number of nodes is shown in Fig.~\ref{fig:scaling-experiments-2} after 50 iterations (200 seconds) and 100 iterations (400 seconds); i.e. each point represents the same wall-clock computation time.  This demonstrates the correctness of MLitB for a given model architecture and learning hyperparameters.  

Due to the data allocation policy that limits the data vector capacity of each node to 3000 vectors, experiments with more nodes process more of the training set during the training procedure.  For example, using only $1$ slave node trains on 3/60 of the full training set.  With 20 nodes, the network is training on the full dataset.  This policy could easily be modified to include data refreshment when running with unallocated data.  

The primary latency issue is due to all clients simultaneously sending gradients to the server at the end of each iteration.  Three simple scaling solutions are 1) increasing the number of master node processes that receive gradients 2) using asynchronous update rules (each slave computes for a random amount of time, then sends updates), reducing the load of any one master node process, and 3) partial communication of gradients (decreasing bandwidth). 

\begin{figure}[b!]
\centering
\includegraphics[width=0.9\columnwidth]{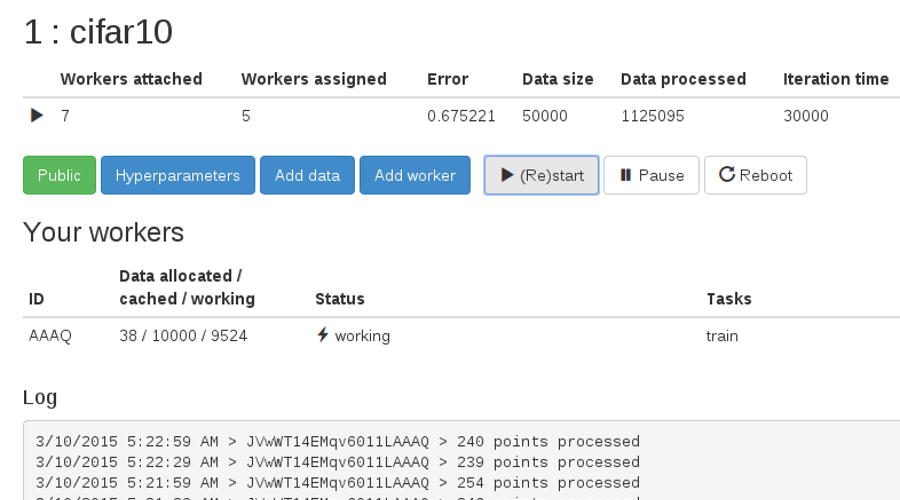}
\caption{{\bf CIFAR-10 project loaded in MLitB.}
}
\label{fig:cifar10-project2}
\end{figure}

\subsection{Walk-through of MLitB Prototype}
We briefly describe how MLitB works from a researcher's point of view.  

\subsubsection*{Specification of Neural Network and Training Parameters}
Using  a minimalist UI (not shown), the researcher can specify their neural network, for example they can add/remove layers of different types, and adjust regularization parameters (L1/L2/dropout) and learning rates.  Alternatively, the researcher can load a previously saved neural network in JSON format (that may or may not have already been trained).  Once a NN is specified (or loaded), it appears in the display, along with other neural networks also managed by the master node.  By selecting a specific neural network, the researcher can then add workers and data (e.g. project {\em cifar10} in Fig.~\ref{fig:cifar10-project2}).  

\subsubsection*{Specification of Training Data}
Image classification data is simple to upload using named directory structures for image labels.  For example, for CIFAR10 all files in the "apple" subdirectory will be given label "apple" once loaded (e.g. the image file   \verb+/cifar10/apple/apple_apple_s_000022.png+).   The entire "cifar10" directory can be zipped and uploaded.  MLitB processes JPEG and PNG formats.  A test set can be uploaded in {\em tracker} mode. 

\subsubsection*{Training Mode}
In the {\em training} mode, a training worker performs as many gradient computations as possible within the iteration duration $T$ (i.e. during the {\em map} step of the main event loop).  The total gradient and the number of gradients is sent to the master, which then in the {\em reduce} step  computes a weighted average of gradients from all workers and takes a gradient step using AdaGrad~\cite{duchi2011adaptive}.  At the end of the main event loop, new neural network weights are sent via Web Sockets to both trainer workers (for the next gradient computations) and to tracker workers (for computing statistics and executing the latest model).  

\begin{figure}[t!]
\centering
\includegraphics[width=\columnwidth]{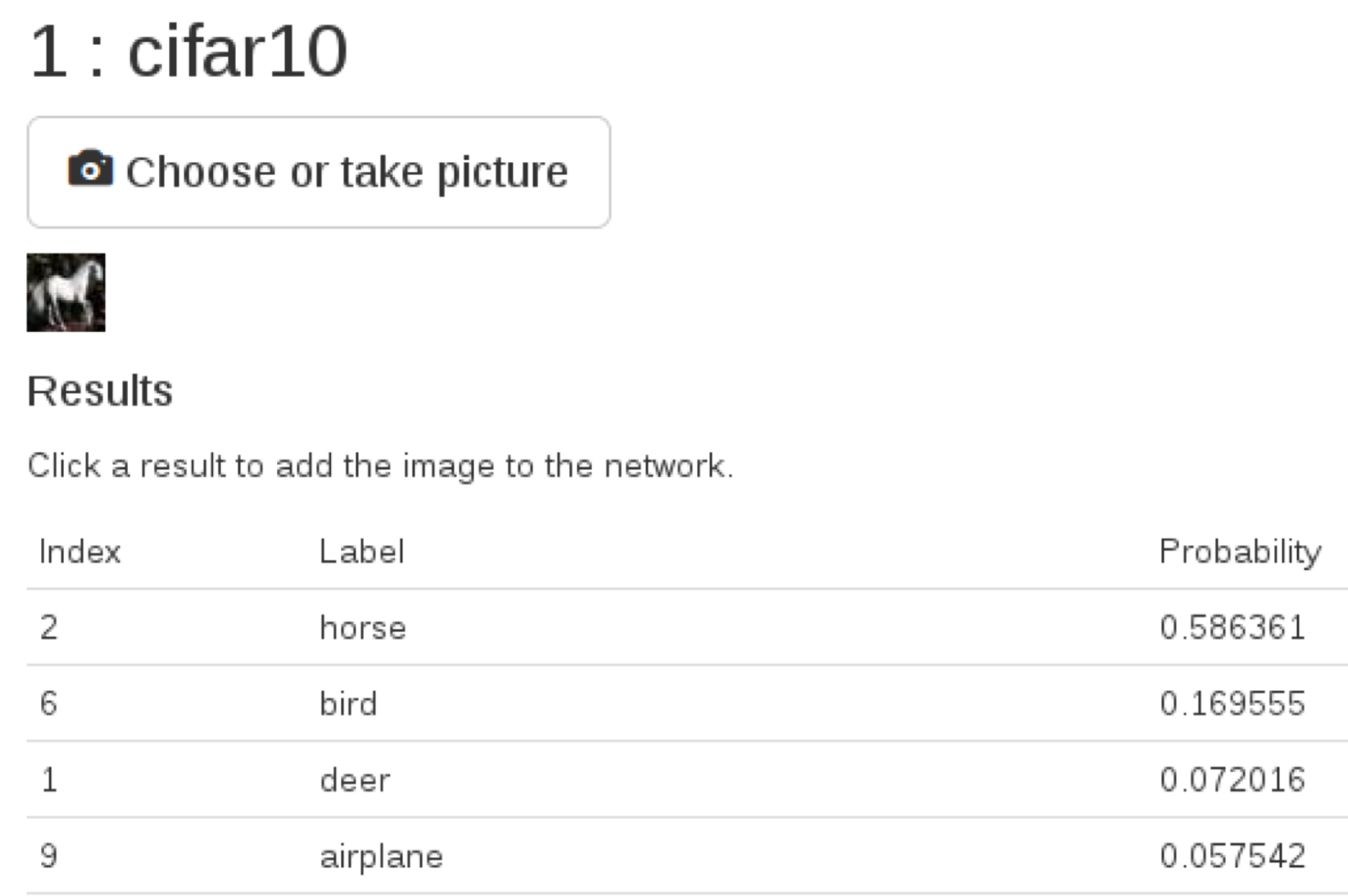}
\caption{{\bf Tracking model (model execution)}: The label of a test image is predicted using the latest NN parameters.  Users can execute a NN prediction using an image stored on their device or using their device's camera.  In this example, an image of a horse is correctly predicted with probability $0.687$ (the class-conditional predictive probability).
}
\label{fig:cifar10-classify_horse}
\end{figure}

\begin{figure}[t!]
\centering
\includegraphics[width=0.9\columnwidth]{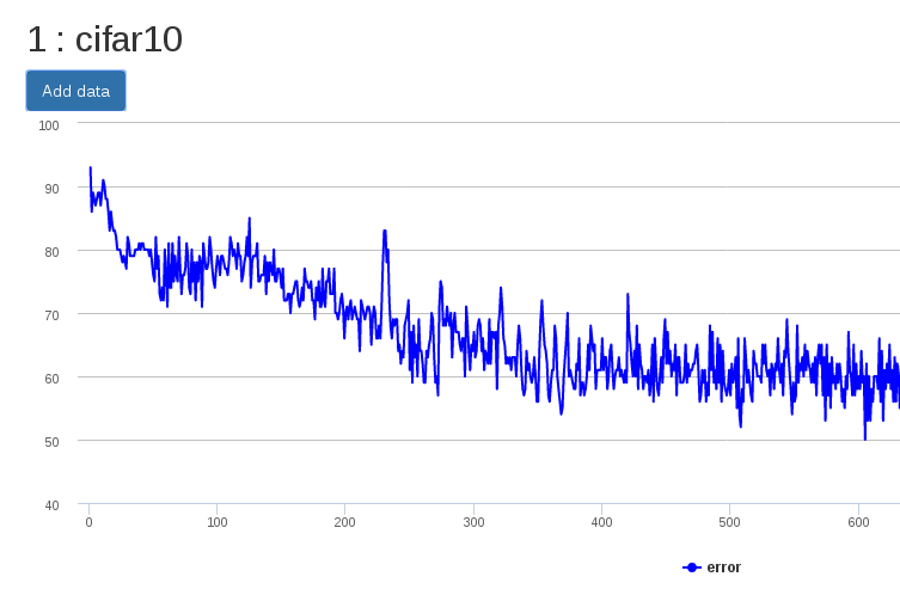}
\caption{{\bf Tracking mode (classification error).}  A test dataset can be loaded and its classification error rate tracked over iterations; here using a NN trained on CIFAR-10.
}
\label{fig:cifar10-test100_2}
\end{figure}
\subsubsection*{Tracking Mode}
There are two possible functions in {\em tracking} mode: 1) executing the neural network on test data, and 2) monitoring classification error on an independent data set.  For 1, users can predict class labels for images taken with a device's camera or locally stored images.  Users can also learn a new classification problem on the fly by taking a picture and giving it a new label; this is treated as a new data vector and a new output neuron is added dynamically to the neural network if the label is also new.  Fig.~\ref{fig:cifar10-classify_horse} shows a test image being classified by the {\em cifar10} trained neural network.  For 2, users create a statistics worker and can upload test images and track their error over time; after each complete evaluation of the test images, the latest neural network received from the master is used.  Fig.~\ref{fig:cifar10-test100_2} shows the error for {\em cifar10} using a small test set for the first 600 parameter updates.   

\subsubsection*{Archiving Trained Neural Network Model}
The prototype does not include a research closure specification.  However, it does provide  easy archiving functionality.  At any moment, users can download the entire model specification and current parameter values in JSON format.  Users can then share or initialize a new training session with the JSON object by uploading it during the model specification phase, which represents a high-level of reproducibility.  Although the JSON object fully specifies the model, it does not include training or testing code.  Despite this shortcoming,  using a standard protocol is simple way of providing a lightweight archiving system.
%

\subsection{Limitations of MLitB Prototype}\label{sec:limitations}
In this section we briefly discuss the limitations of the current prototype; later in  Section~\ref{sec:challenges} we will discuss the challenges we face in scaling MLitB to a massive level. 

Our scaling experiment demonstrates that the MLitB prototype can accommodate up to 64 clients before latency significantly degrades its performance.  Latency, however, is primarily affected by the length of an iteration and by size of the neural network.  For longer iterations, latency will become a smaller portion of the main event loop.  For very large neural networks, latency will increase due to bandwidth pressure. 

As discussed previously, the main computational efficiency loss is due to the synchronization requirement of the master event loop.  This requirement causes the master server to be idle while the clients are computing and the clients to wait while the master processes all the gradients.  As the size of the full gradients can be large (at least $>1MB$ for small neural networks), the network bandwidth is quickly saturated at the end of a computation iteration and during the parameter broadcast.  By changing to an asynchronous model, the master can continuously process gradients and the bandwidth can be maximally utilized.   By communicating partial gradients, further efficiency can be attained.  We leave this for future work.

There is a theoretical limit of 500MB data storage per client (the viable memory of a web-browser).  In our experience, the practical limit is closer to 100MB at which point performance is lost due to memory management issues.  We found that 1MB/sec bandwidth was achievable on a local network, which meant that it could handle images on MNIST and CIFAR-10 easily, but would stall for larger images.  With respect to Deep Neural Networks, the data processing ability of a single node was limited (especially is one compared to sophisticated GPU enables libraries~\cite{bastien2012theano}). Although we were most interested in the scaling performance, we note that naive convolution implementations significantly slow performance.  We found that reasonable sized images, up to $100\times100\times3$ pixels, can be processed on mobile devices in less than a second without convolutions, but can take several seconds with convolutions, limiting its usefulness.  In the future, near native or better implementations will be required for the convolutional layers.



\section{Related Work}\label{sec:related}
MLitB has been influenced by a several different technologies and ideas presented by previous authors and from work in different specialization areas.  We briefly summarize this related work below.

\subsection{Volunteer Computing}

BOINC~\cite{anderson2004boinc} is an open-source software library used to set up a grid computing network,  allowing  anyone with a desktop computer connected to the internet to participate in computation; this is called {\em public resource computing}.   Public resource or volunteer computing was popularized by  SETI@Home~\cite{anderson2002seti}, a research project that  analyzes radio signals from space in the search of signs of extraterrestrial intelligence.  More recently, protein folding has emerged as significant success story~\cite{lane2013milliseconds}.   Hadoop~\cite{shvachko2010hadoop} is an open-source software system for storing very large datasets and executing user application tasks on large networks of computers. MapReduce~\cite{dean2008mapreduce} is a general solution for performing computation on large datasets using  computer clusters.


\subsection{JavaScript Applications}

In~\cite{cushing2013distributed} a network of distributed web-browsers called {\em WeevilScout} is used for complex computation (regular expression matching and binary tree modifications) using a JavaScript engine.  It uses similar technology (Web Workers and Web Sockets) as MLitB.  
%
ConvNetJS~\cite{karpathy2014convnetjs} is a JavaScript implementation of a convolutional neural-network, developed primarily for educational purposes, which is capable of building diverse neural networks to run in a single web browser and trained using  stochastic gradient descent; it can be seen as the non-distributed predecessor of MLitB.

\subsection{Distributed Machine Learning}
The most performant deep neural network models are trained with sophisticated scientific libraries written for GPUs~\cite{bergstra+al:2010-scipy-theano, jia2014caffe, collobert2011torch7} that provide orders of magnitude computational speed-ups compared to CPUs.  Each implements some form of stochastic gradient descent  (SGD)~\cite{bottou2010large} 
 as the training algorithm.   Most implementations are limited to running on the cores of a single machine and  by extension the memory limitations of the GPU.  Exceptionally, there are distributed deep learning algorithms that use a farm of GPUs (e.g. {\em Downpour SGD}~\cite{dean2012large}) and  farms of commodity servers (e.g. {\em COTS-HPS}~\cite{coates2013deep}).  Other distributed ML algorithm research includes the parameter server model~\cite{li2014scaling}, parallelized SGD~\cite{zinkevich2010parallelized}, and distributed SGD~\cite{ahn2014distributed}.  MLitB could potentially push commodity computing to the extreme using pre-existing devices, some of which may be GPU capable, with and without an organization's existing computing infrastructure.  As we discuss below, there are still many open research questions and opportunities for distributed ML algorithm research.

\section{Opportunities and Challenges}\label{sec:challenges}

In tandem with our vision, there are several directions the next version of MLitB can take, both in terms of the library itself and the potential kinds of applications a ubiquitous ML framework like MLitB can offer.  We first focus on the engineering and research challenges we have discovered
during the development of our prototype, along with some we expect as the project grows.  Second, we look at the opportunities MLitB provides, not only based on the research directions the challenges uncovered, but also novel application areas that are perfect fits for MLitB.  In Section~\ref{sec:future} we preview the next concrete steps in MLitB development.


\subsection{Challenges}
We have identified three keys engineering and research challenges that must be overcome for MLitB to achieve its vision of learning models a global scale.    
%
%

\subsubsection*{Memory Limitations} 
State-of-the-art Neural Network models have huge numbers of parameters.  This prevents them from fitting onto mobile devices.
 There are two possible solutions to this problem.  The first solution is to learn or use smaller neural networks.  Smaller NN models have shown promise on image classification performance, in particular the Network in Network~\cite{lin2013network} model from the Caffe model zoo, is 16MB, and outperforms AlexNet which is 256MB~\cite{jia2014caffe}.  It is also possible to first train a deep neural network then use it to train a much smaller, shallow neural network~\cite{ba2014deep}.
 Another solution is to distribute the NN (during training and prediction) across clients.  An example of this approach is Downpour SGD~\cite{dean2012large}. 

\subsubsection*{Communication Overhead} 
With large models, large of numbers of parameters are communicated regularly.  This is a similar issue to memory limitation and could benefit from the same solutions.  However, given a fixed bandwidth  and asynchronous parameter updates, we can ask what parameter updates (from master to client) and which gradients (from client to master) should be communicated.  
 An algorithm could transmit a random subset of the weight gradients, or send the most informative.  
 In other words, given a fixed bandwidth budget, we want to maximize the information transferred per iteration.  

\subsubsection*{Performance Efficiency} Perhaps the biggest argument against scientific computing with JavaScript is its computation performance.  We disagree that this should prevent the widespread adoption of browser-based, scientific computing because the goal of several groups to achieve native performance in JavaScript~\cite{jsv8,jsasm} and GPU kernels  are becoming part of existing web engines (e.g. WebCL\footnote{WebCL by Kronos: \url{www.khronos.org/webcl}.}) and they can be seamlessly incorporated into existing JavaScript libraries, though they have yet to be written for ML.  


\subsection{Opportunities}

\subsubsection*{Massively Distributed Learning Algorithms} 
The challenges just presented are obvious areas of future distributed machine learning research (and are currently being developed for the next version of MLitB).  Perhaps more interesting is, at a higher level, that the MLitB vision raises novel questions about what it means to train models on a global scale.  For instance, what does it mean for a model to be trained across a global internet of heterogeneous and unreliable devices?  Is there a single model or a continuum of models that are consistent locally, but different from one region to another?  How should a model adapt over long periods of time?  
These are largely untapped research areas for ML.

%

\subsubsection*{Field Research} 
Moving data collection and predictive models onto mobile devices makes is easy to bring models into the field.   Connecting users with mobile devices to powerful NN models can aid field research by bringing the predictive models to the field, e.g. for fast labeling and data gathering.  For example, a pilot program of crop surveillance in Uganda currently uses bespoke computer vision models for detecting pestilence (insect eggs, leaf diseases, etc)~\cite{quinn2011modeling}.  Projects like these could leverage publicly available, state-of-the-art computer vision models to bootstrap their field research.

\subsubsection*{Privacy Preserving Computing and Mobile Health} 
Our MLitB framework provides a natural platform for the development of real privacy-preserving application~\cite{dwork2008differential} by naturally protecting user information contained on mobile devices, yet allowing the data to be used for valuable model development.   The current version of MLitB does not provide privacy preserving algorithms such as~\cite{han2010privacy}, but these could be easily incorporated into MLitB.  It would therefore be possible for a collection of personal devices to collaboratively train machine learning models using sensitive data stored locally and with modified training algorithms  that guarantee privacy. 
%
  One could imagine, for example, using privately stored images of a skin disease to build a classifier based on large collection of disease exemplars, yet with the data always kept on each patient's mobile device, thus never shared, and trained using privacy preserving algorithms. 

\subsubsection*{Green Computing}
One of our main objectives was to provide simple, cheap, distributed computing capability with MLitB.   Because MLitB runs with minimal software installation (in most cases requiring none), it is possible to use this framework for low-power consumption distributed computing.  By using existing organizational resources running in low-energy states (dormant or near dormant) MLitB can wake the machines, perform some computing cycles, and return them to their low-energy states.  This is in stark contrast to a data center approach which has near constant, heavy energy usage~\cite{nrdcreport}.

\section{Future MLitB Development}\label{sec:future}
The next phases of development will focus on the following directions: a visual programming user interface for model configuration, development of a library of ML models and algorithms, development of performant scientific libraries in JavaScript with and without GPUs, and model archiving with the development of a research closure specification.

\subsection{Visual Programming} 
Many ML models are constructed as chains of processing modules.  This lends itself to a visual programming paradigm, where the chains can be constructed by dragging and dropping modules together.  This way models can be visualized and compared, dissected, etc.  Algorithms are tightly coupled to the model and a visual representation of the model can allow interaction with the algorithm as it proceeds.  For example, learning rates for each layer of a neural network can be adjusted while monitoring error rates (even turned off for certain layers), or training modules can be added to improve learning of hidden layers for very deep neural networks, as done in~\cite{SzegedyLJSRAEVR14}.  With a visual UI it would be easy to pull in other existing, pre-trained models, remove parts, and train on new data.  For example, a researcher could start with a pre-trained image classifier, remove the last layer, and easily train a new image classifier, taking advantage of an existing, generalized image representation model.

\subsection{Machine Learning Library}
We currently have built a prototype around an existing JavaScript implementation of DNNs~\cite{karpathy2014convnetjs}.  In the near future we plan on implementing other models (e.g. latent Dirichlet allocation) and algorithms (e.g. distributed MCMC~\cite{ahn2014distributed}).  MLitB is agnostic to learning algorithms and therefore is a great platform for researching novel distributed learning algorithms.  To do this, however, MLitB will need to completely separate machine learning model components from the MLitB network.  At the moment, the prototype is closely tied to its neural network use-case.  Once separated, it will be possible for external modules to be added by the open-source community.
%
 
\subsection{GPU implementations}
Implementation of GPU kernels can bring MLitB performance up to the level of current state-of-the-art scientific libraries such as Theano~\cite{bergstra+al:2010-scipy-theano,bastien2012theano} and Caffe~\cite{jia2014caffe}, while retaining the advantages of using heterogeneous devices.  For example, balancing computational loads during training is very simple in MLitB and any learning algorithm can be shared by GPU powered desktops and mobile devices.  Smart phones could be part of the distributed computing process by permitting the training algorithms to use short bursts of GPU power for their calculations, and therefore limiting battery drain and user disruption.

\subsection{Design of Research closures}
MLitB can save and load JSON model configurations and parameters, allowing researchers to share and build upon other researchers' work.  However, it does not quite achieve our goal of a research closure where all aspects---code, configuration, parameters, etc--are saved into a single object.  In addition to research closures, we hope to develop a model zoo, akin to Caffe's for posting and sharing research.  Finally, some kind of system for verifying models, like recomputation.org, would further strengthen the case for MLitB being truly reproducible (and provide backwards compatibility).

\section{Conclusion}
In this paper we have introduced MLitB: Machine Learning in the Browser, an alternative framework for ML research based entirely on using the browser as the computational engine.  The MLitB {\bf vision} is based upon the overarching objectives that provide {\em ubiquitous ML} capability to every computing device, {\em cheap distributed computing}, and {\em reproducible research}.  
The MLitB {\bf prototype} is written entirely in JavaScript and makes extensive use of existing JavaScript libraries, including Node.js for servers, Web Workers for non-blocking computation, and Web Sockets for communication between clients and servers.  We demonstrated the potential of MLitB on a ML use-case:  Deep Neural Networks trained with distributed Stochastic Gradient Descent using heterogenous devices, including dedicated grid-computing resources and mobile devices, using the same interface and with no client-side software installation.  Clients simply connect to the server and computing begins.  This use-case has provided valuable information for future versions of MLitB, exposing both existing challenges and interesting research and application opportunities.  We have also advocated for a framework which supports reproducible research; MLitB naturally provides this by allowing models and parameters to be saved to a single object which can be reloaded and used by other researchers  immediately.

\ifCLASSOPTIONcompsoc
  \section*{Acknowledgments}
\else
  \section*{Acknowledgment}
\fi

The authors acknowledge funding support from Amsterdam Data Science and computing resources from SurfSara.

\ifCLASSOPTIONcaptionsoff
  \newpage
\fi



\bibliographystyle{IEEEtran}
%
\bibliography{mlitb}

%


%
%
%
%




\end{document}